\documentclass[useAMS]{mn2e}
\usepackage{epsfig}

\def\lsim{\,\lower2truept\hbox{${<\atop\hbox{\raise4truept\hbox{$\sim$}}}$}\,}
\def\gsim{\,\lower2truept\hbox{${> \atop\hbox{\raise4truept\hbox{$\sim$}}}$}\,}
\title[Low-frequency evolution of radio sources]{A model for the cosmological evolution
of low frequency radio sources}
\author[Massardi et al.]{
\parbox[t]{\textwidth}
{Marcella Massardi$^{1}$\thanks{E-mail: marcella.massardi@oapd.inaf.it},
Anna Bonaldi$^1$, Mattia Negrello$^2$, Sara Ricciardi$^3$, Alvise Raccanelli$^4$, Gianfranco De Zotti$^{1,5}$}
\vspace*{8pt} \\
$^{1}$ INAF, Osservatorio Astronomico di Padova, Vicolo dell'Osservatorio 5, I-35122 Padova,
        Italy\\
$^{2}$ Department of Physics and Astronomy, Open University, Walton Hall, Milton Keynes MK7 6AA,
        United Kingdom \\
$^{3}$ INAF/IASF, Sezione di Bologna, Via Gobetti, 101, I-40129 Bologna, Italy \\
$^{4}$ Institute of Cosmology \& Gravitation, University of Portsmouth, Portsmouth P01 3FX, UK \\
$^{5}$ SISSA, Via Beirut 2--4, I-34014 Trieste, Italy }

\begin{document}

\maketitle

%%%%%%%%%%%%%%%%%%%%%%%%%%%%%%%%%%%%%%%%%%%%%%%%%%%%%%%%%%%%%%%%%%%%%%%%%%
\begin{abstract}
We present a new evolutionary model that describes the population properties of radio sources at frequencies $\lsim 5$ GHz, thus complementing the De Zotti et al. (2005) model, holding at higher frequencies. We find that simple analytic luminosity evolution is still sufficient to fit the wealth of available data on local luminosity functions, multi-frequency source counts, and redshift distributions. However, the fit requires a luminosity-dependent decline of source luminosities at high redshifts, at least for steep-spectrum sources, thus confirming earlier indications of a ``downsizing'' also for radio sources.  The upturn of source counts at sub-mJy levels is accounted for by a straightforward extrapolation, using the empirical far-IR/radio correlation, of evolutionary models matching the far-IR counts and redshift distributions of star-forming galaxies. We also discuss the implications of the new model for the interpretation of data on large-scale clustering of radio sources and on the Integrated Sachs-Wolfe (ISW) effect, and for the investigation of the contribution of discrete sources to the extragalactic background. As for the ISW effect, a new analysis exploiting a very clean CMB map, yields at a substantially higher significance than reported before.
\end{abstract}

\begin{keywords}
galaxies: active --- galaxies: evolution --- radio continuum: general.
\end{keywords}

%%%%%%%%%%%%%%%%%%%%%%%%%%%%%%%%%%%%%%%%%%%%%%%%%%%%%%%%%%%%%%%%%%%%%%%%%%
\section{Introduction}\label{sec:Intro}

The still most widely used evolutionary models for radio sources at $\nu \lsim 5\,$GHz date back to the 1990's (Dunlop \& Peacock 1990; Toffolatti et al. 1998). Although these models proved to be very useful benchmarks even today, several of their building blocks need to be updated, starting from the underlying cosmology (a matter dominated flat universe with $H_0=50\,\hbox{km}\,\hbox{s}^{-1}\,\hbox{Mpc}^{-1}$). Even more importantly, large amounts of new data have been accumulating since then, including new deep/large area surveys (see De Zotti et al. 2009 for a review and references), identifications of sub-mJy sources at 1.4 GHz (Seymour et al. 2008; Bondi et al. 2008; Ibar et al. 2009; and references therein), accurate determinations of the local luminosity function (Magliocchetti et al. 2002; Condon et al. 2002; Best et al. 2005; Mauch \& Sadler 2007), redshift distributions for complete samples (Best, Rottgering \& Lehnert 1999; Willott et al. 2002; Jackson et al. 2002; Brookes et al. 2008; Gendre \& Wall 2008). Complete redshift information on source samples at different flux limits is especially critical to clarify the evolutionary properties. For example, as shown by Brookes et al. (2008), the observed redshift distribution of the CENSORS sample, complete down to $S_{1.4\rm GHz}=7.2\,$mJy, is not well reproduced by any of the Dunlop \& Peacock's (1990) models. All that points to the need of updated models.

Many key scientific issues in this field are still open: what is the nature of evolution? is it luminosity dependent? is there a clear evidence for a decline of the comoving space density of radio sources at high redshifts? are flat- and steep-spectrum sources obeying the same evolution laws, as expected in the framework of `unification' schemes? which sources are responsible for the upturn of differential source counts at sub-mJy levels?

In this paper we will investigate to what extent the new data allow us to progress towards answering these questions. The analysis is similar and complementary to the one by De Zotti et al. (2005), holding for $\nu \ge 5\,$GHz. In fact, manageable models cannot fully take into account the complex and multiform spectral behaviour of radio sources. They must adopt schematic descriptions, that most frequently boil down to considering two populations, steep- and flat-spectrum sources, each with a single power-law spectrum, $S\propto \nu^{-\alpha}$. At best, a Gaussian distribution of spectral indices is allowed for. In contrast, real source spectra, especially those classified as ``flat'', are generally not power-laws, but have complex shapes, showing spectral bumps, flattenings or inversions (i.e. flux increasing with increasing frequency), frequently bending to steeper power-laws at higher frequencies. This means that power-law approximations can only hold for limited frequency ranges. In particular, the De Zotti et al. (2005) model becomes increasingly inaccurate with decreasing frequency below 5 GHz. Furthermore, the high frequency data sets analyzed by De Zotti et al. (2005) are dominated by compact, flat-spectrum sources and do not strongly constrain the evolutionary properties of the steep-spectrum population. Much stronger constraints are provided by the very abundant data collected at low frequencies, which are the subject of the present analysis.

In \S\,\ref{sec:Model} we present our parametrization of the evolving luminosity functions of AGN-powered radio sources and its application to estimate source counts and redshift distributions. In \S\,\ref{sec:lowlum} we deal with the contributions of star-forming galaxies to the source counts. In \S\,\ref{sec:backgr} the integrated emission of sources is compared to observational estimates of the radio background.  In \S\,\ref{sec:largescale} we discuss the implications of the new model for the interpretation of the angular correlation function of radio sources and of data on the Integrated Sachs--Wolfe effect. Finally, in \S\,\ref{sec:Conclusions} we summarize our conclusions.

We have adopted a flat $\Lambda$CDM cosmology with $\Lambda=0.7$ and ${\rm H}_0 = 70\, {\rm km\, s^{-1}\, Mpc^{-1}}$.

%%%%%%%%%%%%%%%%%%%%%%%%%%%%%%%%%%%%%%%%%%%%%%%%%%%%%%%%%%%%%%%%%%%%%%%%%%
\section{AGN powered sources}\label{sec:Model}
%%%%%%%%%%%%%%%%%%%%%%%%%%%%%%%%%%%%%%%%%%%%%%%%%%%%%%%%%%%%%%%%%%%%%%%%%%

\subsection{Data sets}
\subsubsection{Local luminosity functions}
\begin{figure*}
\begin{center}
\includegraphics[width=12cm, angle=90]{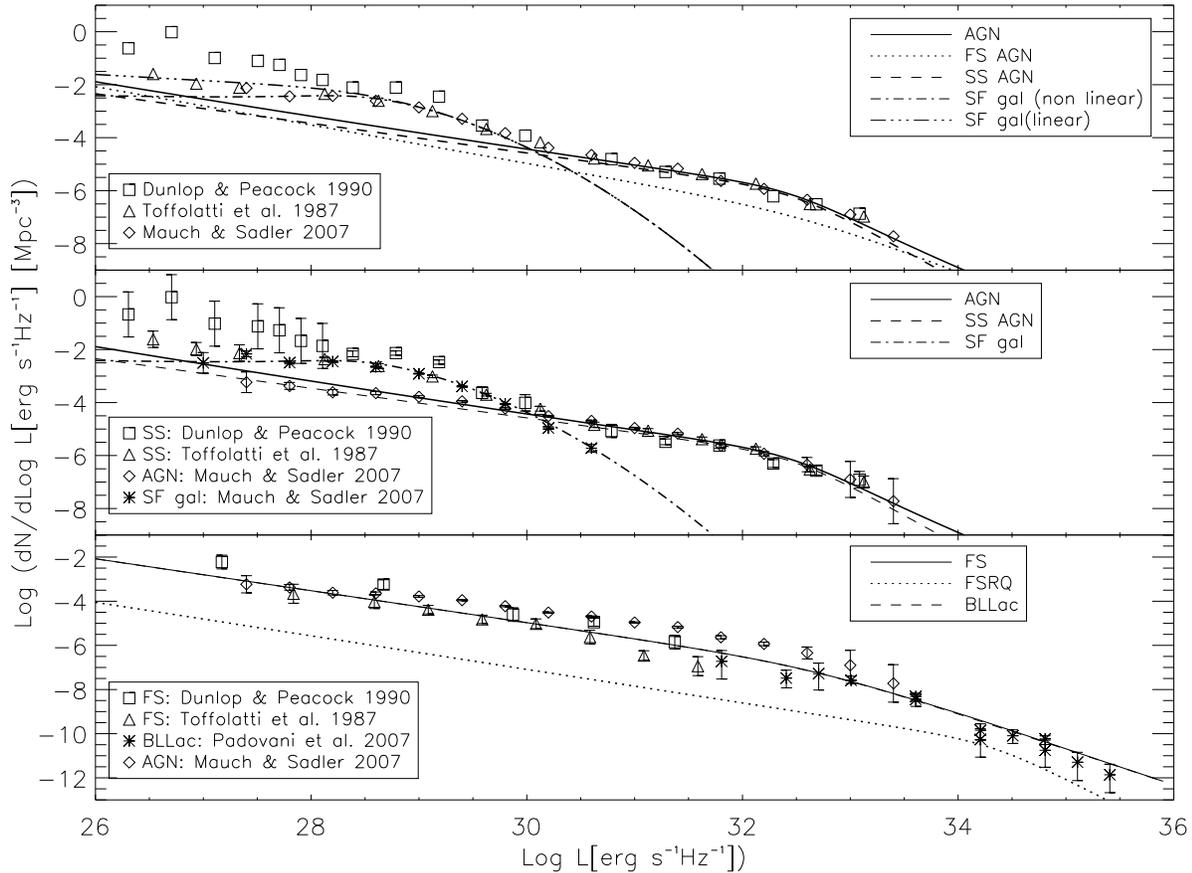}
\caption{Comparison of model with observed local radio luminosity functions (LRLFs) at 1.4 GHz. The {\it upper panel} refers to the global (flat- plus steep-spectrum source, FS and SS respectively) luminosity function. The solid line shows the best fit LRLF of radio AGNs and the dotted and the dashed lines give the contributions of flat- and steep-spectrum components, respectively. The (triple) dot-dashed line shows the LRLF of star-forming (SF) galaxies, obtained by (linearly) non-linearly extrapolating the $60\,\mu$m luminosity function by means of the well established far-IR/radio correlation (see \S \ref{sec:lowlum}). The estimate by Toffolatti et al. (1987; triangles) is in remarkably good agreement with the the recent assessment by Mauch \& Sadler (2007; diamonds), over the whole luminosity range. The estimate by Dunlop \& Peacock (1990, their ``determination 1''; squares) agrees well with for $\log L(\hbox{erg}\,\hbox{s}^{-1}\,\hbox{Hz}^{-1})\gsim 29.5$, but is systematically higher at lower luminosities. Error bars are omitted to avoid overcrowding.
The {\it central panel} refers to steep-spectrum sources. The diamonds and the asterisks show, respectively, the LRLFs of radio AGNs and of star-forming galaxies derived by Mauch \& Sadler (2007) who did not distinguish between steep- and flat-spectrum sources. The triangles and the squares show the estimates by Toffolatti et al. (1987) and Dunlop \& Peacock (1990, ``determination 1'') for steep spectrum sources only, but including both AGNs and star-forming galaxies. The solid line is the best fit LF of steep-spectrum AGNs and the dot-dashed line that of star-forming galaxies. The latter is the result of the non-linear extrapolation of the local $60\,\mu$m luminosity function (see text).
Finally, the {\it bottom panel} shows the LRLF of flat-spectrum sources (AGNs only). The observational estimates are, again, from Toffolatti et al. (1987) and Dunlop \& Peacock (1990). At the highest luminosities these join smoothly with the LRLF of BL Lacs by Padovani et al. (2007; asterisks). At the lowest luminosities the LRLF of Dunlop \& Peacock (1990, ``determination 1''; squares) is slightly high in comparison with both Toffolatti et al. (triangles) and the global (flat- plus steep-spectrum) LRLF of AGNs  by Mauch \& Sadler (2007; diamonds), shown for comparison. The dashed and the dotted lines are the contributions of BL Lacs and of Flat Spectrum Radio Quasars (FSRQs), respectively, to the best fit LRLF of flat-spectrum sources (solid line). The Dunlop \& Peacock (1990) and Toffolatti et al. (1987) data have been rescaled to the value of $H_0$ used in this paper and extrapolated from 2.7 to 1.4 GHz using a spectral index of 0.8 for steep-spectrum sources, and of 0.1 for flat-spectrum sources. } \label{fig:lumfunc}
\end{center}
\end{figure*}
A spectacular progress towards the determination of the local radio luminosity function (LRLF) at 1.4 GHz has been achieved combining large-area spectroscopic surveys (Las Campanas, SDSS, 2dF and 6dF) with the NVSS and FIRST surveys (Machalski \& Godlowski 2000; Magliocchetti et al. 2002; Sadler et al. 2002; Best et al. 2005; Mauch \& Sadler 2007). Condon et al. (2002) estimated the LRLF by cross-matching NVSS sources brighter than 2.5 mJy with galaxies brighter than $m_p = 14.5\,$mag in the Uppsala Galaxy Catalogue (UGC; Nilson 1973).

We will mostly refer to the analysis by Mauch \& Sadler (2007) who derived the LRLF from a large, well-defined sub-sample of radio sources with $S_{1.4\rm GHz}\ge 2.8\,$mJy from the NVSS catalogue, identified with galaxies observed in the Second Incremental Data Release (Jones et al. 2005) of the 6 degree Field Galaxy Survey (6dFGS). It may be noted, however, that the NVSS is only 50\% complete at 2.5 mJy, and its completeness rises rapidly to 99\% at 3.4 mJy (Condon et al. 1998). The incompleteness is therefore still quite significant ($\sim 30\%$) at 2.8 mJy. This likely implies that the true uncertainties are larger than the quoted statistical errors. In fact, estimates of the local luminosity functions using the samples with complete (or almost complete) redshift information (see \S\,\ref{sec:zdistr}), and in particular the steep-spectrum sub-sample of K\"uhr et al. (1981) and the  CoNFIG sample (Gendre \& Wall (2008) give, on average, local space densities about 0.2 dex lower in the luminosity range for which these samples have sufficient statistics (around $\log L_{1.4\,{\rm GHz}}(\hbox{erg}\,\hbox{s}^{-1}\,\hbox{Hz}^{-1}) \simeq 32$; see also Rigby et al. 2008). In our analysis we have therefore adopted an error of at least 0.2 dex
%$\delta(dN/d\log L_{1.4\,{\rm GHz}}(\hbox{erg}\,\hbox{s}^{-1}\,\hbox{Hz}^{-1}))=0.2$
for the Mauch \& Sadler (2007) LRLF of radio AGNs. Also, Mauch \& Sadler (2007) have included in their ``local'' sample sources up to $z \simeq 0.2$. However, the $V/V_{\rm max}$ test for the AGN sub-sample showed evidence for cosmological evolution at more than $3\sigma$. We have therefore restricted ourselves to sources at $z \leq 0.05$, i.e. to AGNs with $\log L_{1.4\,{\rm GHz}}({\rm erg\ s^{-1} Hz^{-1}})<33.3$, for which evolution is expected to have a negligible effect.

A further important recent advance is the derivation of separate luminosity functions for the star-forming galaxies and for the AGNs (see the upper panel of Fig.~\ref{fig:lumfunc}). This has allowed us to easily deal with the different evolutionary properties of the two source populations. Note that star-forming galaxies have, in most cases, steep spectra.

On the other hand, recent studies did not attempt to disentangle the contributions to the LRLF of sources of different spectral classes. The time-honoured determinations of the LRLFs of flat- and steep-spectrum sources by Dunlop \& Peacock (1990) and Toffolatti et al. (1987) are compared, in the top panel of Fig.~\ref{fig:lumfunc}, with the total LRLF by Mauch \& Sadler (2007). The agreement with the Toffolatti et al. (1987) estimates is extremely good, while the estimate by Dunlop \& Peacock (1990) is somewhat high at low luminosities. As shown in the central panel of the figure, for $\log L[{\rm erg\ s^{-1} Hz^{-1}}]\lsim 30$, the global LRLF of steep-spectrum sources closely matches Mauch \& Sadler's (2007) determination of the starburst LRLF, while at higher luminosities it matches the AGN LRLF. The bottom panel shows that the Toffolatti et al. (1987) LRLF of flat-spectrum sources joins smoothly with the estimated LRLF of BL Lac objects obtained by Padovani et al. (2007). We have included the Toffolatti et al. (1987) LRLFs of flat- and steep-spectrum sources and the Padovani et al. (2007) LRLF for BL Lacs in our data-base.

\begin{figure*}
\begin{center}
\includegraphics[width=12cm, angle=90]{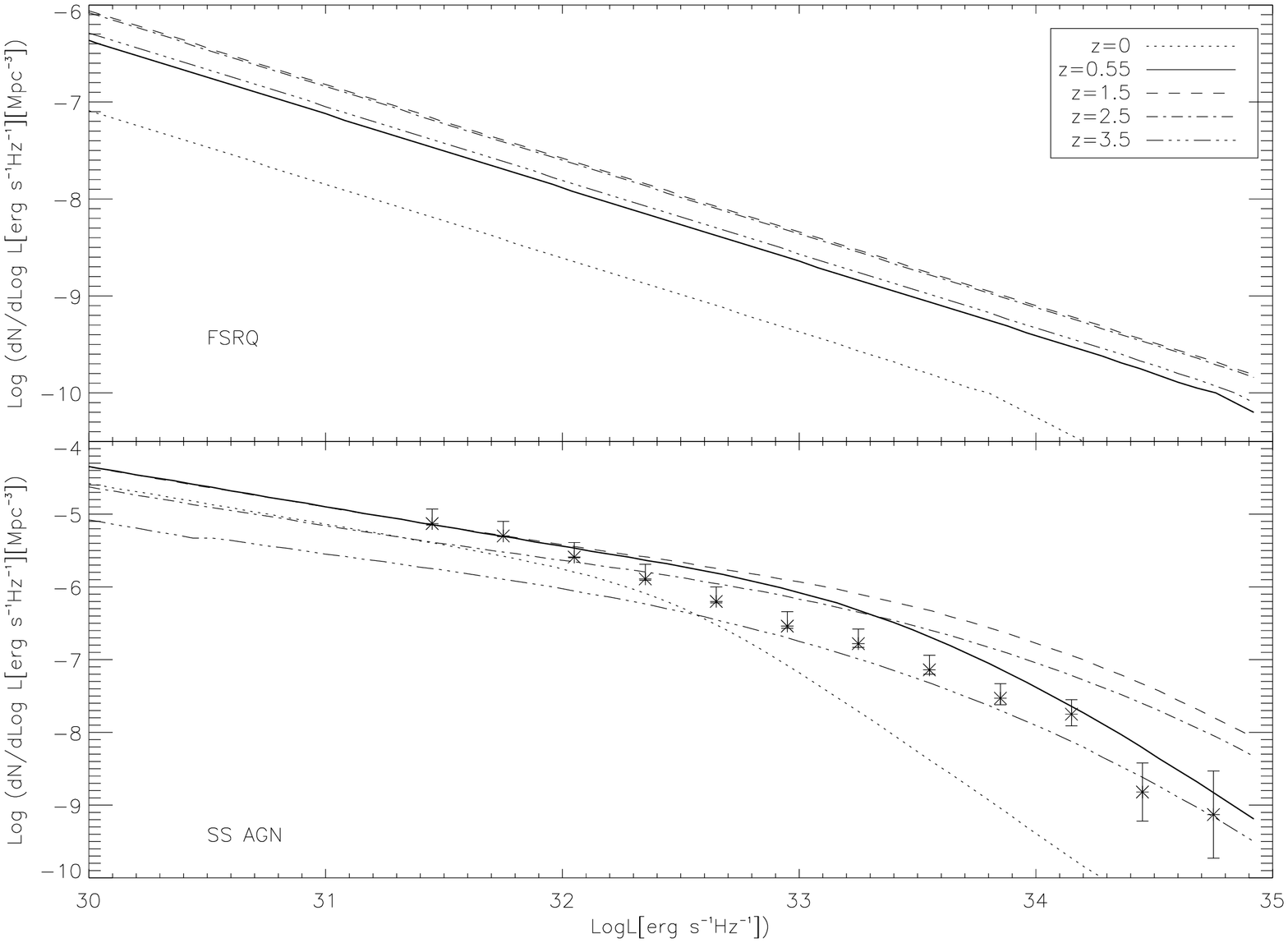}
\caption{Model 1.4 GHz RLF at different redshifts, specified in the inset, for FSRQs (upper panel) and steep-spectrum (SS) AGNs (lower panel). The Donoso et al. (2009) 1.4 GHz RLF at $z \sim 0.55$ is also shown (asterisks) for comparison (the model RLF at the same redshift is given by the solid line). This data does not distinguish between flat and steep-spectrum objects, but at 1.4 GHz and at $z \sim 0.55$ the sample is expected to be mostly constituted by steep-spectrum sources; it is therefore compared with the model for SS AGNs. } \label{fig:flum_z}
\end{center}
\end{figure*}

Donoso et al. (2009) presented a determination of the 1.4 GHz radio luminosity function at $z \sim 0.55$ by cross correlating the NVSS and FIRST catalogs with the MegaZ Luminous Red Galaxy (MegaZ-LRG) catalog derived from Sloan Digital Sky Survey imaging data. The combined catalog has an estimated reliability of $\simeq 98.3$ per cent and completeness level of about 95 per cent. Accurate photometric redshifts are available for all the objects. Although the sample is very large, and the statistical errors on the luminosity function correspondingly small, it is important to note that the radio luminosity function misses radio galaxies falling outside the MegaZ-LRG colour selection criteria, as well as radio-loud quasars. On the whole, the Donoso et al. estimate may be too low by 0.1--0.2 dex (see Fig. \ref{fig:flum_z}). For this reason we have adopted a positive error of at least 0.2 dex for this dataset.

%$\delta(dN/d\log L_{1.4\,{\rm GHz}}(\hbox{erg}\,\hbox{s}^{-1}\,\hbox{Hz}^{-1}))=0.2$ for this dataset.
%

\subsubsection{Source counts}\label{sect:counts}
\begin{figure*}
\begin{center}
\includegraphics[width=12cm, angle=90]{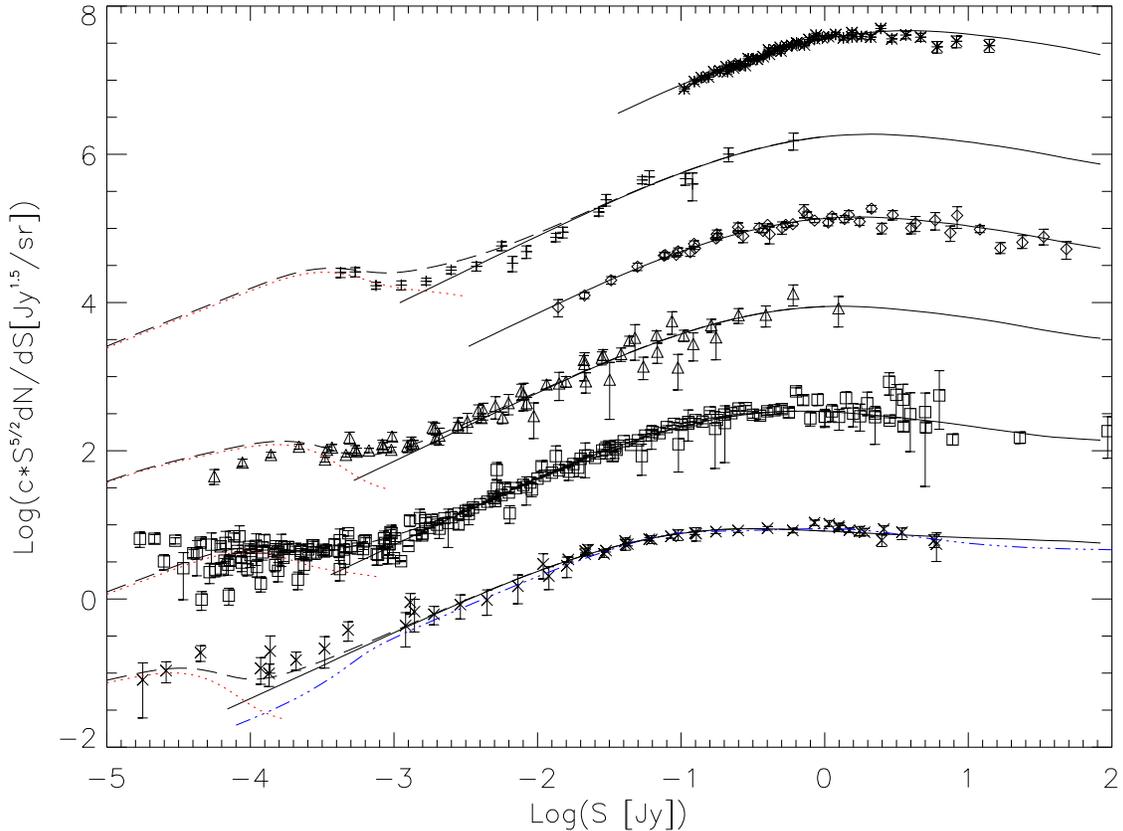}
\caption{Comparison of our best fit model with the observed source counts (multiplied by a factor $c$ for a better visualization) at several frequencies. From top to bottom: 150 MHz ($c=10000$), 325 MHz ($c=1000$), 408 MHz ($c=100$), 610 MHz ($c=10$), 1.4 GHz ($c=1$), 5.0 GHz ($c=0.1$). The solid and the dotted lines show our model counts for AGNs and starforming galaxies, respectively, while the dashed lines show the sum of the two contributions. The dot-dashed line at 5 GHz shows the counts yielded by the De Zotti et al. (2005) model for AGNs; they are essentially indistinguishable from those of the present model, except at sub-mJy levels. For a complete list of references and tabulations of the data points see De Zotti et al. (2009).} \label{fig:srccntall}
\end{center}
\end{figure*}
\begin{figure*}
\begin{center}
\includegraphics[width=12cm, angle=90]{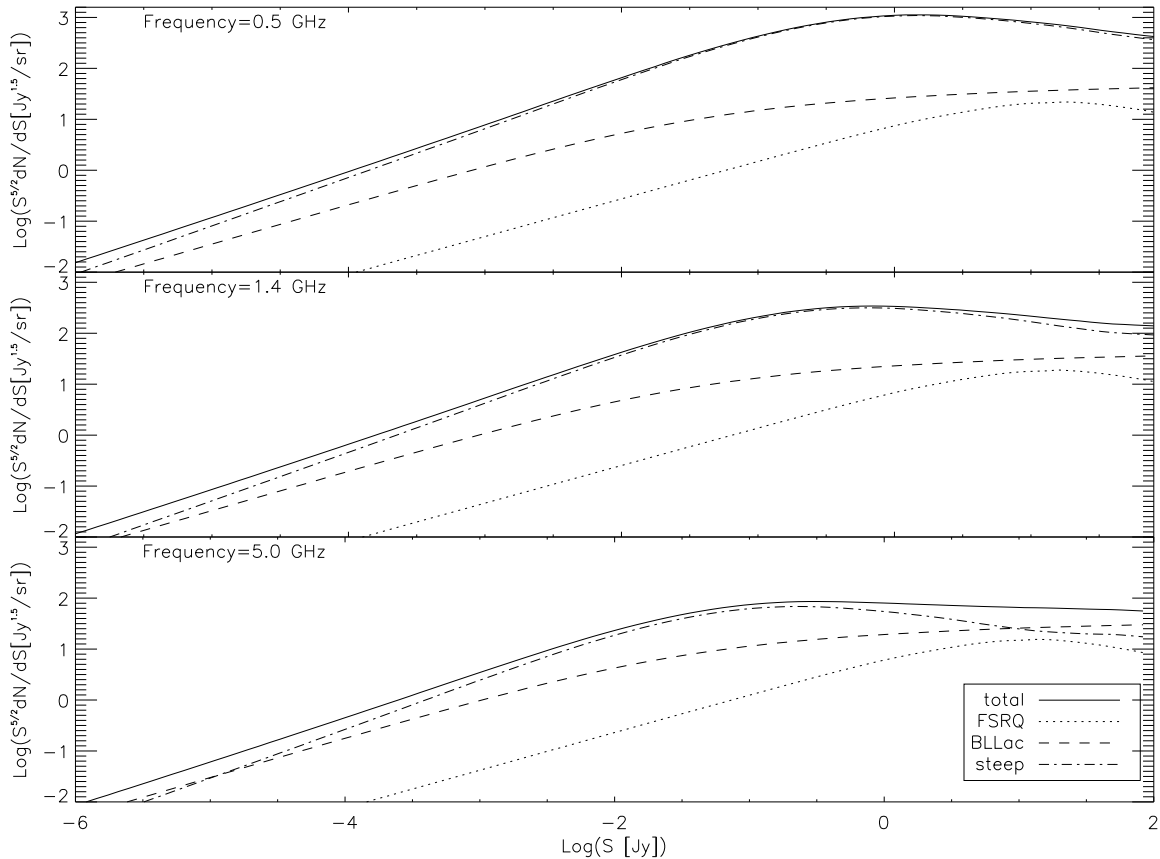}
\caption{Contributions of different source populations (see inset) to the source counts at 0.5, 1.4, and 5.0 GHz. } \label{fig:modelcounts}
\end{center}
\end{figure*}
An updated compilation of radio source counts is presented by De Zotti et al. (2009)\footnote{Tables of the source counts and model results can be downloaded from the webpage http://web.oapd.inaf.it/rstools/srccnt/srccnt\_tables.html.}. Our analysis deals with those in the frequency range $0.15\leq \nu \leq 5\,$GHz. While above $\sim 1\,$mJy counts are dominated by AGNs, the contribution of star-forming galaxies becomes increasingly important at fainter flux levels (Windhorst et al. 1985; Benn et al. 1993; Hopkins et al. 1998; Seymour et al. 2004; Muxlow et al. 2005; Moss et al. 2007). Therefore sub-mJy source counts were not taken into account in our model fitting process, unless the contributions of radio AGNs were reliably singled out. This is the case for Seymour et al. (2008) who used several diagnostics to determine the source counts of AGNs down to less than 0.1 mJy at 1.4 GHz; their faint AGN counts were included in the dataset used for the fitting. As shown in \S\,\ref{sec:lowlum}, the contribution to the counts of star-forming galaxies can be obtained by straightforwardly extrapolating models fitting mid- and far-IR data, without introducing any adjustable parameter. On the bright side, an issue of concern is the possibility that giant radio sources, which have global flux density above the selection limit of a sample are missed because of resolution effects (Riley 1989). The observed counts may therefore be lower limits to the true counts.

In the case of small-area surveys, the sampling variance due to source clustering can be a further significant addition to the uncertainties. The total variance of the counts is (Peebles 1980):
\begin{equation}
\left\langle {n-\langle n \rangle \over \langle n }\rangle \right\rangle^2 = {1\over \langle n \rangle} + \sigma_v^2
\end{equation}
with
\begin{equation}\label{eq:sigmav}
\sigma_v^2={1\over \Omega^2}\int \int w(\theta)\,d\Omega_1\,d\Omega_2
\end{equation}
where $\theta$ is the angle between the solid angle elements $d\Omega_1$ and $d\Omega_2$, and the integrals are over the solid angle covered by the survey. The angular correlation function of NVSS and FIRST sources is consistent with a power-law shape (Blake \& Wall 2002; Overzier et al. 2003):
\begin{equation}\label{eq:w}
w(\theta) \simeq 10^{-3} (\theta/\hbox{deg})^{-0.8},
\end{equation}
for angular separations up to at least $4^\circ$.  Inserting eq.~(\ref{eq:w}) in eq.~(\ref{eq:sigmav}) we get
\begin{equation}
\sigma^2=2.36\times 10^{-3}  (\Omega/\hbox{deg}^2)^{-0.4}.
\end{equation}
This contribution was summed in quadrature to the Poisson errors for surveys over areas $\le 25\,\hbox{deg}^2$. However  the actual differences between published source counts from independent fields are much larger than those accounted for by Poisson plus clustering fluctuations (see Fig.~4 of Condon 2007). As discussed by Condon (2007) the dominant contributions to such differences are likely due to calibration errors, resolution corrections, corrections for primary-beam attenuation, applied in different ways by different observers. For example, the deepest surveys cover a single primary-beam area with nonuniform sensitivity, so the counts contain large corrections for primary gain; and surveys made with high angular resolution must be corrected for missing extended sources. This means that using the Poisson plus clustering errors to weight the source count data in the comparison with the model may be misleading. On the other hand, it is difficult to find consistent alternative weighting schemes, and this leaves some level of discretion in the comparison of the model to the data.

%%%%%%%%%%%%%%%%%%%%%%%%%%%%%%%%%%%%%%%%%%%%%%%%%%%%%%%%%%%%%%%%%%%%%%%%%%
\subsubsection{Redshift distributions}\label{sec:zdistr}
\begin{figure*}
\begin{center}
\includegraphics[width=10cm, height=16cm, angle=90]{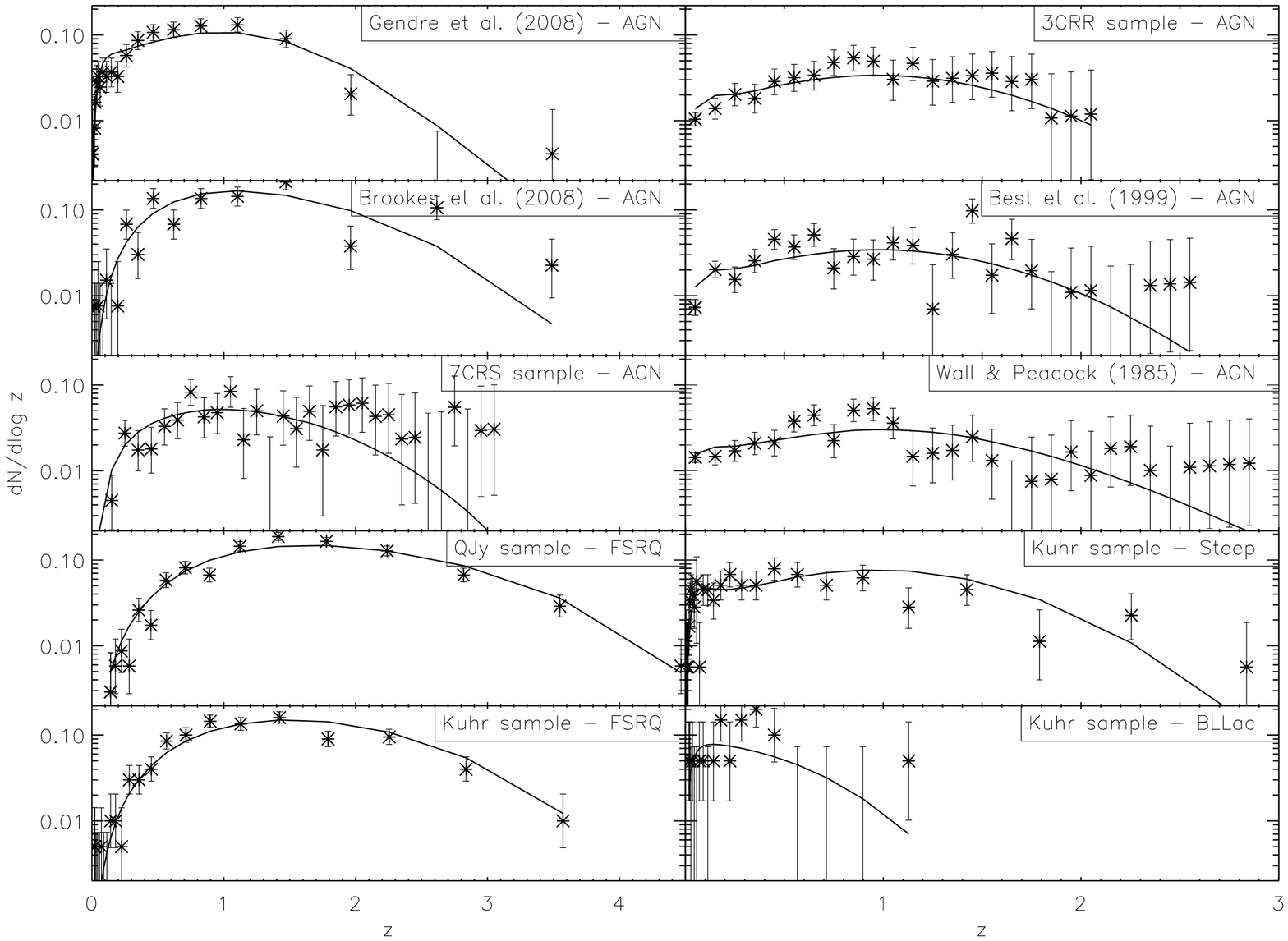}
\vskip10pt
\caption{Normalized redshift distributions yielded by our best fitting model (solid lines) compared with data by  Gendre \& Wall (2008), 3CRR  (Laing, Riley \& Longair 1983), Brookes et al. (2008), Best, R{\"o}ttgering, \& Lehnert (1999), Wall \& Peacock (1985), 7CRS (Willott et al. 2002; Lacy, Bunker \& Ridgway 2000), Jackson et al. (2002; quarter-Jy, QJy, sample, which includes only FSRQs) and the redshift distributions for the steep-spectrum, FSRQ, and BL Lac sub-samples of the K\"uhr et al. (1981) sample; see insets.} \label{fig:allzdist}
\end{center}
\end{figure*}

As mentioned in \S\,\ref{sec:Intro}, particularly important new constraints on evolutionary models have been provided by the redshift distribution for the Combined EIS-NVSS Survey Of Radio Sources (CENSORS; Brookes et al. 2008). We have considered the complete sample of 136 AGNs with $S_{\rm lim, 1.4\,GHz}=7.2\,$mJy over a $6\ {\rm deg}^2$ area. The 9 sources having a lower limit for the redshift have been taken into account using the Kaplan-Meier estimator [routine KMESTM in the software package ASURV Rev 1.2 (Isobe \& Feigelson, 1990), which implements the survival analysis methods presented in Feigelson \& Nelson (1985) and Isobe, Feigelson \& Nelson (1986)]. Poisson errors have been adopted.

While the CENSORS sample is unique in providing information on the redshift distribution of faint sources, spectroscopically complete (or close to completeness) samples of bright sources include:

\begin{itemize}
\item the 3CRR sample (Laing, Riley \& Longair 1983) containing a total of 173 sources, complete for
$S_{178\rm MHz} > 10.9\,$Jy and with spectral index information and redshifts for all sources %\footnote{Data available at 3crr.extragalactic.info/ or www.science.uottawa.ca/~cwillott/3crr/3crr.html}
\item the Best, R{\"o}ttgering, \& Lehnert (1999) sample of 178 sources brighter than $S_{408\rm MHz} = 5\,$Jy, with complete redshift information (Best et al. 2003)
\item the Wall \& Peacock (1985) 2-Jy sample of 239 sources selected at 2.7 GHz, only 5 of which are missing a redshift estimate
\item the 7C redshift survey (7CRS; Willott et al. 2002; Lacy, Bunker \& Ridgway 2000): 103 sources with $S_{151\rm MHz} \gsim 0.5\,$Jy and full redshift information.
\item the Combined NVSS-FIRST Galaxies (CoNFIG) sample (Gendre \& Wall 2008) comprising 274 sources with
$S_{1.4\rm GHz} > 1.3\,$Jy.  Spectroscopic redshifts are available for 230 sources (84\%) and photometric estimates
were made for 14 sources, bringing the total redshift coverage to 89\%.
\end{itemize}

\noindent While all these samples do not distinguish between steep- and flat-spectrum sources, the K\"uhr et al. (1981) catalog provides also spectral information. The sample comprises 518 sources to a 5 GHz flux density limit of 1 Jy, over an area of 9.811 sr. Based on the catalogued spectral indices, 299 sources are flat-spectrum ($\alpha < 0.5$, $S \propto \nu^{-\alpha}$) and 219 are steep-spectrum ($\alpha \ge 0.5$). The former population includes 212 flat-spectrum radio quasars (FSRQs), 200 of which (94\%) have measured redshift, 26 BL Lacs (20 of which, 77\%, with measured redshift, and 61 either classified as galaxies or missing a morphological classification. We have distributed the latter sources among FSRQs and BL Lacs in proportion of the fraction of classified sources. The fraction of steep-spectrum sources with measured redshift is of 81\%.

Jackson et al. (2002) presented 99\% complete optical identifications of the 878 sources of the Parkes 2.7 GHz Quarter-Jansky flat-spectrum sample. Redshifts are available for 58\% of the sources. Following De Zotti et al. (2005) we have defined a complete sub-sample aiming at maximizing the fraction of sources with measured redshift. The sub-sample includes the sources brighter than 0.25 Jy detected in areas surveyed at least down to this limit at $\delta >-50^\circ$, for a total of 514 objects of which 370 are FSRQs (93\% with redshift), 47 are BL Lacs (21\% with redshift), 95 are flat-spectrum radio galaxies, and 2 are unidentified. Only for FSRQs a meaningful redshift distribution can be derived.

\subsection{The model}\label{sec:model}
\begin{figure*}
\begin{center}
\includegraphics[width=9cm, height=17cm, angle=90]{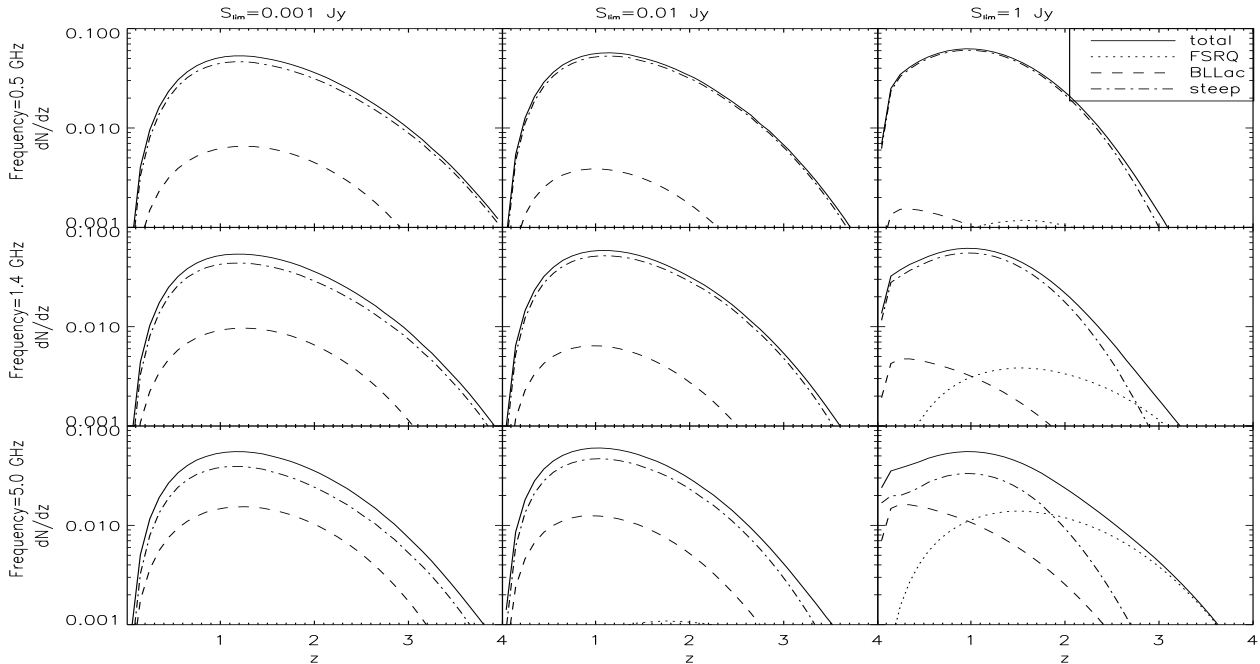}
\caption{Contributions of the different populations to the redshift distributions yielded by our best fitting model at different frequencies (0.5, 1.4, 5.0 GHz, from top to the bottom) and different flux limits (0.001, 0.01, 1 Jy from left to right). } \label{fig:modelzdist}
\end{center}
\end{figure*}

Our model includes two flat-spectrum populations with different evolutionary properties (FSRQs, and BL Lacs) and a single steep-spectrum population. For each population we adopt a simple power-law spectrum, $S\propto \nu^{-\alpha}$, with $\alpha_{\rm FSRQ}=\alpha_{\rm BLLac}=0.1$, and $\alpha_{\rm steep} = 0.8$.

The epoch-dependent {\it comoving} luminosity functions (in units of ${\rm Mpc^{-3}\, (d\log L)^{-1}}$) are modeled as double power-laws:
\begin{equation}
\Phi(L, z)=\frac{n_0}{(L/L_\star)^a+(L/L_\star)^b}.
\end{equation}
The high-$z$ evolution of radio sources has been the subject of a hot, still not completely settled, debate (Dunlop \& Peacock 1990; Shaver et al. 1996; Jarvis \& Rawlings 2000; Wall et al. 2005; Cruz et al. 2007). Standard luminosity evolution models ($L_{\star}(z)= L_{\star}(0) \exp[k_{\rm evo}\tau(z)]$, $\tau(z)$ being the lookback time in units of the Hubble time, $H_0^{-1}$), without a high-$z$ decline of the comoving luminosity function, did not yield any acceptable fit to the data on steep-spectrum AGNs and on FSRQs. For these source populations we have therefore adopted a more complex luminosity evolution law:
\begin{equation}\label{eq:evol}
L_{\star}(z)\!=\! L_{\star}(0)\,{\rm dex}\!{\left[k_{\rm evo}z\!\left(\!2z_{\rm top}\!-\!{z^{m_{\rm ev}}z_{\rm top}^{(1-m_{\rm ev})}}\!/(1\!+\!m_{\rm ev})\right)\right]},
\end{equation}
where $z_{\rm top}$ is the redshift at which $L_{\star}(z)/ L_{\star}(0)$ reaches its maximum and $|m_{\rm ev}|\le 1$.

Although this improved substantially the quality of the fit, significant discrepancies between the model and the data remained. In particular, the model fell short at reproducing the highest $z$ points of the redshift distributions of the Brookes et al. (2008) sample and the K\"uhr et al. (1981) steep-spectrum sub-sample (see Fig. \ref{fig:allzdist}). We can force the model to fit those points, but at the cost of visibly worsening the fit to the counts and/or to the local luminosity functions. A better global fit was achieved allowing $z_{\rm top}$ to vary with luminosity as:
\begin{equation}\label{eq:ztop}
z_{\rm top}=z_{{\rm top},0}  + {\delta z_{\rm top}\over 1+ L_{\star}(0)/L},
\end{equation}
$z_{{\rm top},0} $ and $\delta z_{\rm top}$ being free parameters. The fit gave positive values for both of them, consistent with earlier indications that the high-$z$ decline of the space density is more pronounced and starts at lower redshifts for less powerful sources (Hook et al. 1998; Waddington et al. 2001; Vigotti et al. 2003; Cirasuolo et al. 2005, 2006; Wall et al. 2009 in preparation). The luminosity dependence of the high-$z$ decline is qualitatively similar to the {\it downsizing} observed for galaxies and optically and X-ray selected quasars (Cowie et al. 1996; Barger et al. 2005; P\'erez-Gonz\'alez et al. 2008).

Actually, as shown by Table~\ref{tab:bestfit}, the luminosity dependence of the peak redshift required by the data is substantial for the steep-spectrum population but is weak ($\delta z_{\rm top}\ll 1$) for FSRQs. This is expected since the contributions to the observables (source counts and redshift distributions) of moderate to low luminosity flat-spectrum sources are dominated by BL Lacs, so that the evolution of low luminosity FSRQs is poorly constrained.

In the case of BL Lacs we have adopted an evolutionary form similar to those used for the FSRQ and steep-spectrum sources, but, since the data are not enough to constrain several evolutionary parameters for this population, we have used the simpler form of eq.~(\ref{eq:evol}), with $m_{\rm ev}=1$ (fixed) and a luminosity-independent redshift cut-off.

\begin{table}
 \caption{Best fitting model parameters. The spectral indices were fixed at $\alpha_{\rm FSRQ}=\alpha_{\rm BLLac}=0.1$, and $\alpha_{\rm steep} = 0.8$}
  \label{tab:bestfit}
  \begin{tabular}{lccc}
    \hline
    Parameters          & FSRQ  & BLLac & SS-AGNs   \\
    \hline
    a                   & 0.760 & 0.723 & 0.559     \\
    b                   & 2.508 & 1.618 & 2.261     \\
    $\log n_0$          &-10.382& -6.879& -5.970    \\
    $\log L_\star(0)$   & 34.323& 32.638& 32.490    \\
    $k_{\rm evo}$       & -0.996& 0.208 & 1.226     \\
    $z_{{\rm top},0}$   & 1.882 & 1.282 & 0.977     \\
    $\delta z_{\rm top}$& 0.018 &  -    & 0.842     \\
    $m_{\rm ev}$        & -0.166&  1    & 0.262     \\
    \hline
  \end{tabular}
\end{table}

\subsection{Multidimensional minimization algorithms}

\begin{table}
 \caption{Contributions to $\chi^2$ of each data set for the best fit values of the parameters (see Table~\protect\ref{tab:bestfit}).}
  \label{tab:chis}
  \begin{tabular}{lccc}
    \hline
    Data set          & $\chi^2$  & N$_{\rm data}$ & $\chi^2/$N$_{\rm data}$  \\
    \hline
150 MHz	counts [1]  &	110.0	&	45		&	2.444	\\
325 MHz counts [1]  &	52.2	&	17		&	3.071	\\
408 MHz counts [1]  &	57.7	&	44		&	1.311	\\
610 MHz counts [1]  &	63.8	&	37		&	1.724	\\
1.4 GHz counts [1]  &	363.0	&	115		&	3.156	\\
4.8 GHz counts [1]  &	50.7	&	46		&	1.102	\\
$z$ distr. [2]      &	29.1	&	20		&	1.455	\\
$z$ distr. [3] FSRQ	&	24.9	&	16		&	1.556	\\
$z$ distr. [4] FSRQ	&	9.5	&	24		&	0.396	\\
$z$ distr. [4] BLLac&	12.5	&	19		&	0.658	\\
$z$ distr. [4] SS    &	36.8	&	35		&	1.051	\\
$z$ distr. [5]	     &	23.0	&	26		&	0.885	\\
$z$ distr. [6]	     &	11.8	&	21		&	0.561	\\
$z$ distr. [7]	     &	26.7	&	41		&	0.651	\\
$z$ distr. [8]	     &	32.6	&	26		&	1.254	\\
$z$ distr. [9]	     &	22.2	&	29		&	0.766	\\
LLF [10] AGN	       &	7.4	&	12		&	0.617	\\
LLF [11] SS	     &	13.6	&	5		&	2.720	\\
LLF [12] BLLac	     &	33.7	&	12		&	2.808	\\
LF [13] at $z=0.55$	&	4.4	&	8		&	0.553	\\
\hline
Total	                &	963.0	&	598	&	1.610	\\
\hline
 \end{tabular}
 {\small{\\References: [1] De Zotti et al. (2009); [2] Brookes et al. (2008); [3] Jackson et al. (2002); [4] K\"uhr et al. (1981); [5] Gendre \& Wall (2008); [6] Laing, Riley \& Longair (1983); [7] Willott et al. (2002);[8] Best, R{\"o}ttgering, \& Lehnert (1999); [9] Wall \& Peacock (1985); [10] Mauch \& Sadler (2007); [11] Toffolatti et al. (1987); [12] Padovani et al. (2007); [13] Donoso et al. (2009)}}
\end{table}
The best-fit model has been found by minimizing the sum of $\chi^2$ values obtained comparing the model with data on luminosity functions, redshift distributions and source counts. The minimization was first performed using the \emph{multi-dimensional downhill simplex method} (exploiting the AMOEBA routine, Press et al. 1992). The method is computationally fast even in the case of a large number of parameters. However, it does not allow us to bracket a solution, but, starting from an initial guess, modifies the parameters defining its path towards the closest minimum. So, if the parameter space has several local minima, the method can be trapped at a secondary minimum or can find a minimum for physically implausible values of some parameters.

In order to check the sensitivity of the downhill simplex method to the initial guess and to confirm the final results, we have also exploited the \emph{simulated annealing} minimization technique (Aarts \& Korst 1989). This method exploits an analogy between the way in which a metal cools and freezes into a minimum energy crystalline structure and the search for a minimum in a more general system. The major advantage of simulated annealing over other methods is its ability to avoid becoming trapped at secondary minima. Also it allows us to specify the plausible range for each parameter. The algorithm employs a random search which not only accepts changes that decrease the merit function, but also some changes that increase it. The probability of the latter to occur is inversely proportional to the increase in the merit function. The disadvantage is that it is computationally demanding, especially for a large number of parameters, as in our case. Therefore, we have used it only to check the stability of the results obtained with the other method.

Care must be taken to avoid that some data sets influence too heavily the results. In particular the redshift distributions, which contain far fewer points than the counts, and have larger error bars may be under-weighted. To check whether this may seriously affect the results we have repeated the minimization increasing the weights of the redshift distributions by up to a factor of 10. Since the best fit values of the parameters turned out to be remarkably stable we have adopted those obtained without introducing additional weights.

The best fit values of the parameters are listed in Table~\ref{tab:bestfit} and Table~\ref{tab:chis} details the contributions to $\chi^2$ of each data set. The fact that fits of the source counts yield values of $\chi^2/\hbox{N}_{\rm data}$ as high as 2 or even 3 is not really worrisome since large contributions to $\chi^2$ come from data points from different surveys which are inconsistent with other data; as noted in \S\,\ref{sect:counts} differences among source counts from independent fields are much larger than the statistical errors. Allowing for that, the quality of the fit appears to be acceptable for all data sets.

Comparisons of the best fit model with several data sets are shown in Figs.~\ref{fig:lumfunc}, \ref{fig:srccntall}, and \ref{fig:allzdist}. The disaggregation of the counts yielded by the best fit model into the contributions of the different populations is shown in Fig.~\ref{fig:modelcounts}. The contribution of different populations to the redshift distributions for different flux density limits at different frequencies is shown in Fig.~\ref{fig:modelzdist}. The plotted redshift distributions are normalized, i.e. the number of sources in each redshift bin is divided by the total number of sources in the sample. The contribution of star-forming galaxies to the sub-mJy counts will be discussed in the next section.

As mentioned in \S\,\ref{sec:model} we have tried different evolutionary models. Simple standard luminosity evolution models ($L_{\star}(z)= L_{\star}(0) \exp[k_{\rm evo}\tau(z)]$, $\tau(z)$ being the lookback time in units of the Hubble time) produced absolutely unacceptable fits for both steep- and flat-spectrum sources. The fit strongly improved when the evolution function of both steep-spectrum sources and FSRQs included a decline at high-$z$, described by eq.~(\ref{eq:evol}), while keeping simple luminosity evolution for BL Lacs. The global minimum $\chi^2$ became 1140 (to be compared with $\chi^2_{\rm min}=963$ obtained for the model in Table~\ref{tab:bestfit}; see Table~\ref{tab:chis}). A further improvement ($\chi^2_{\rm min}=1060$) was achieved allowing $z_{\rm top}$ to decrease with decreasing luminosity, according to eq.~(\ref{eq:ztop}). Actually, the decrease of $\chi^2_{\rm min}$ was entirely due to data on steep-spectrum sources; setting $\delta z_{\rm top}=0$ for flat-spectrum sources did not change in any appreciable way the quality of the fit. Finally, we used also for BL Lacs the evolutionary law of eq.~(\ref{eq:evol}); this also improved the fit, yielding our final $\chi^2_{\rm min}=963$.

%%%%%%%%%%%%%%%%%%%%%%%%%%%%%%%%%%%%%%%%%%%%%%%%%%%%%%%%%%%%%%%%%%%%%%%%%%
\section{Star-forming galaxies} \label{sec:lowlum}

\begin{figure}
\includegraphics[width=6cm, angle=90]{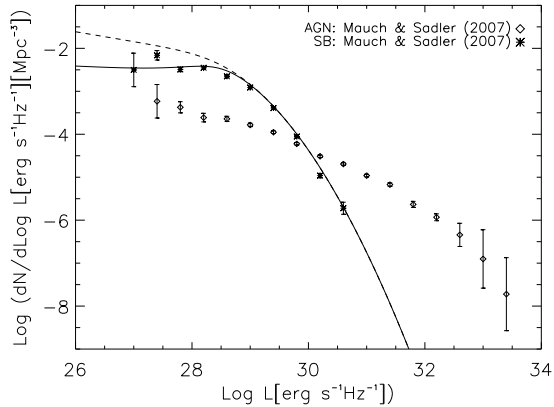}
\caption{Local luminosity functions at 1.4~GHz of radio AGNs (diamonds) and of star-forming galaxies (asterisks), as estimated by Mauch \& Sadler (2007). The lines show the sum of the extrapolations to 1.4 GHz of the $60\mu{\rm m}$ local luminosity functions of ``warm'' (usually interpreted as starburst) and ``cold'' (normal late type) IRAS galaxies by Takeuchi et al. (2003): the dashed line refers to the linear radio/far-IR relationship of eq.~(\ref{eq:corr}), while the solid line is based on that of eq.~(\ref{eq:corr1}), which deviates from linearity at low luminosities. }\label{fig:rlf1d4GHz}
\end{figure}

The radio emission of star-forming galaxies correlates with their star formation rate, as demonstrated by the well-established tight correlation with far-IR emission (Helou et al. 1985; Gavazzi et al. 1986; Condon 1992; Garrett 2002). Yun et al. (2001) found that the overall trend in the range $L_\nu(60\mu{\rm m}) \simeq 10^{30}$--$10^{32.5}\,\hbox{erg}\,\hbox{s}^{-1}\,\hbox{Hz}^{-1}$ is indistinguishable from a linear relation:
\begin{equation}
L_{1.4\rm GHz}=1.16\times 10^{-2}L_\nu(60\mu{\rm m}). \label{eq:corr}
\end{equation}
Galaxies with $L_\nu(60\mu{\rm m})< 10^{30}\,\hbox{erg}\,\hbox{s}^{-1}\,\hbox{Hz}^{-1}$ are found to have radio to far-IR luminosity ratios systematically lower than those given by eq.~(\ref{eq:corr}). The apparent deviation from linearity in the radio/far-IR correlation at low luminosities is supported by a comparison of $60\,\mu$m and 1.4 GHz local luminosity functions (Yun et al. 2001; Best et al. 2005). Simply shifting the $60\,\mu$m luminosity function (Saunders et al. 1990; Takeuchi et al. 2003) along the luminosity axis according to eq.~(\ref{eq:corr}) yields a good match to the radio luminosity function (Best et al. 2005; Mauch \& Sadler 2007, see Fig. \ref{fig:rlf1d4GHz}) for $L_{1.4\rm GHz}\gsim 10^{28.5}\,\hbox{erg}\,\hbox{s}^{-1}\,\hbox{Hz}^{-1}$. At lower luminosities, however, the extrapolated luminosity function lies increasingly above the observed one (Fig.~\ref{fig:rlf1d4GHz}). Full agreement is recovered by replacing eq.~(\ref{eq:corr}) with
\begin{equation}
L_{1.4\rm GHz}  = {1.16\times 10^{-2}L_b \over \left({L_b/L_\nu(60\mu{\rm
m})}\right)^{3.1}
 + {L_b/ L_\nu(60\mu{\rm
m}) }}, \label{eq:corr1}
\end{equation}
in which $L_b = 8.8\times 10^{29}\,\hbox{erg}\,\hbox{s}^{-1}\,\hbox{Hz}^{-1}$.

The tight empirical relationship between radio and far-IR luminosities for star-forming galaxies allows us to take advantage of the wealth of data at far-IR/sub-mm wavelengths to derive the radio evolution properties. We expect a different evolution for starburst and normal late-type galaxies as the starburst activity is likely triggered by interactions and mergers that were more frequent in the past, while in normal galaxies the star-formation rate has probably not changed much over their lifetimes. The bulk of the sub-mm counts measured by SCUBA surveys (Scott et al. 2006; Coppin et al. 2006) is due to yet another population, the sub-mm galaxies (SMGs), interpreted as proto-spheroidal galaxies in the process of forming most of their stars (Granato et al. 2004).

\begin{figure}
\includegraphics[width=6cm, angle=90]{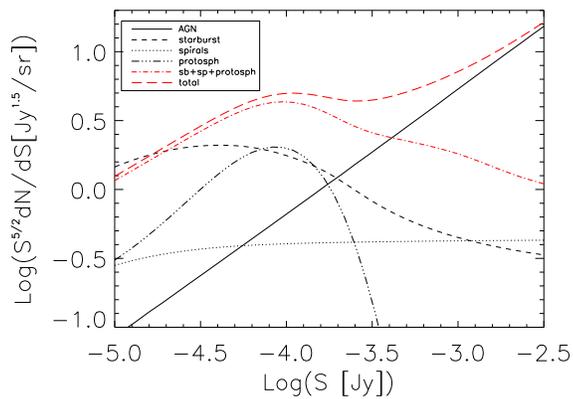}
\caption{Contributions of different populations of star-forming galaxies and of AGNs to the (sub)-mJy source counts at 1.4 GHz. }\label{fig:submJy}
\end{figure}

A straightforward extrapolation to radio frequencies of the evolutionary models by Negrello et al. (2007) for the three populations (normal, starburst and sub-mm galaxies), exploiting eq.~(\ref{eq:corr1}) and the SEDs of NGC$\,6946$ for normal late-type galaxies and of Arp220 for starburst and proto-spheroidal galaxies, yields the counts shown in Fig.~\ref{fig:srccntall}. Adding the contributions of star-forming galaxies to the counts yielded by the model for radio AGNs described in \S\,\ref{sec:model} the sub-mJy counts are nicely reproduced. The counts at tens of $\mu$Jy levels turn out to be dominated by star-forming galaxies. Figure~\ref{fig:submJy} details the contribution of each population of star-forming galaxies to the (sub)mJy counts at 1.4 GHz; the counts of radio AGNs are also shown for comparison.

%%%%%%%%%%%%%%%%%%%%%%%%%%%%%%%%%%%%%%%%%%%%%%%%%%%%%%%%%%%%%%%%%%%%%%%%%%
\section{Contributions to the radio background} \label{sec:backgr}

\begin{figure}
\includegraphics[width=6cm, angle=90]{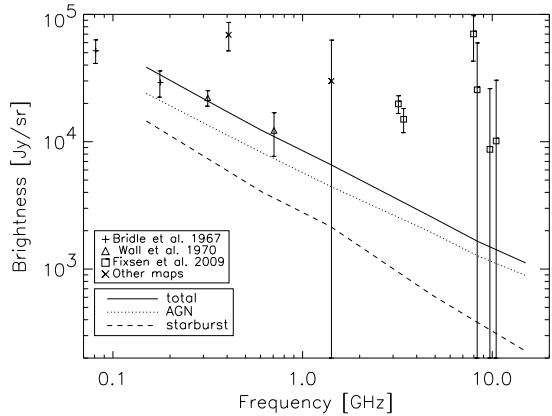}
\caption{Contributions of AGNs and star-forming galaxies to the extragalactic radio background as a function of frequency as yielded by the present model compared with observational estimates. Triangles and crosses refer to estimates based on the T--T plot method (Turtle et al. 1962), while squares refer to estimates by Fixsen et al. (2009) either using ARCADE 2 data (squares) or re-analyzing published data from large area surveys ($\times$).} \label{fig:bck}
\end{figure}

The radio background provides a key constraint on the counts of sources too faint to be individually detected. A summary of the available estimates of the background intensity as a function of frequency is presented in Fig.~\ref{fig:bck}, where the dashed line shows the contribution of extragalactic sources computed using the present model at frequencies of up to 5 GHz and the De Zotti et al. (2005) model at higher frequencies. Our estimates agree very well with those by Gervasi et al. (2008), obtained fitting smooth functions to the observed source counts.

The directly observed counts already account for most of the background intensity estimated by Bridle et al. (1967) and Wall et al. (1970), implying that no much room is left for new source populations and for truly diffuse emissions. The main contribution comes from AGNs. The contribution of star-forming galaxies increases with decreasing frequency, reaching a fraction of 39.5\% percent at 150 MHz.

On the other hand, after subtracting a model for the Galactic emission and the CMB from the measurements of the second-generation balloon-borne experiment ARCADE-2, Fixsen et al. (2009) found excess radiation at 3~GHz about 5 times brighter than the estimated contribution from extragalactic radio sources. From a re-analysis of several large-area surveys at lower frequencies to separate the Galactic and extragalactic components, they derived background intensities much higher than yielded by earlier estimates. It may be noted, however, that their power-law fit to the background spectrum exceeds the minimum {\it total} sky brightness temperature (including Galactic emission) measured by Bridle et al. (1967) at 81.5 MHz.

%%%%%%%%%%%%%%%%%%%%%%%%%%%%%%%%%%%%%%%%%%%%%%%%%%%%%%%%%%%%%%%%%%%%%%%%%%
\section{Implications for the interpretation of data on large-scale structure} \label{sec:largescale}

Extragalactic radio sources are well suited to probe the large-scale structure of the Universe: detectable up to high redshifts, they are unaffected by dust extinction, and can thus provide an unbiased sampling of volumes larger than those probed by optical surveys.
%%%%%%%%%%%%%%%%%%%%%%%%%%%%%%%%%%%%%%%%%%%%%%%%%%%%%%%%%%%%%%%%%%%%%%%%%%

\subsection{The angular correlation function and its implications}
\begin{figure}
\includegraphics[width=9cm]{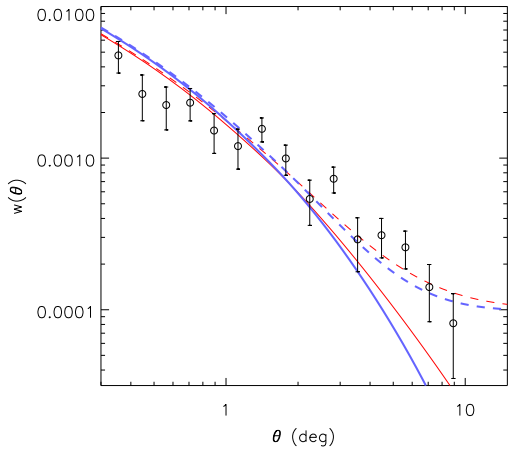}
\caption{Two-point angular correlation function of NVSS sources with $S_{1.4\rm GHz}\ge 10\,$mJy as measured by Blake \& Wall (2002, circles) compared with the model by Negrello et al. (2006, red thin curves) and with the present model (blue thick curves) fitting the redshift distribution by Brookes et al. (2008). The dashed curves include the contribution of a constant offset $\epsilon=0.0001$ to $w(\theta)$ in order to account for the effect of possible spurious density gradients in the survey.}
\label{fig:wth}
\end{figure}
Thanks to the deep radio surveys covering large areas of the sky, such as the NVSS (Condon et al. 1998), the SUMSS-MGPS (Mauch et al. 2003, Murphy et al. 2007), the FIRST (Becker et al. 1995), the WENSS (Rengelink et al. 1997), the angular correlation function, $w(\theta)$, of radio sources has been measured with high statistical significance up to large angular scales (Cress et al. 1996;  Magliocchetti et al. 1998; Rengelink 1999; Blake \& Wall 2002; Overzier et al. 2003; Blake et al. 2004), well beyond those reached by optical surveys (Padmanabhan et al. 2009). Amongst the cited surveys, the NVSS is characterized by the most extensive sky coverage and can thus provide the best clustering statistics. The $w(\theta)$ measured for NVSS sources brighter than 10 mJy, is well described by a power-law of slope $-0.8$ extending from $\sim 0.1$ degrees up to scales of almost 10 degrees (Blake \& Wall 2002).

The angular two-point correlation function of a given class of objects is related to the spatial correlation function, $\xi(r,z)$, by Limber's (1953) equation:
\begin{equation}
w(\theta)=\!\int dz{\mathcal
N}^{2}(z) \!\int \! d(\delta z) \xi[r(\delta
z,\theta),z] \!\left/\left[\int \! dz{\mathcal
N}(z)\right]^{2}\right.
\label{eq:wth_obj}
\end{equation}
where ${\mathcal N}(z)$ is the redshift distribution and $r(\delta z,\theta)$ is the {\it comoving} spatial distance between two objects located at redshifts $z$ and $z+\delta z$ and separated by an angle $\theta$ on the sky. On sufficiently large scales, where the clustering signal is produced by galaxies residing in distinct dark-matter halos, and under the assumption of a one-to-one correspondence between sources and their host halos, $\xi(r,z)$ can be written as the product of the correlation function of dark matter, $\xi_{\rm DM}$, times the square of the bias parameter, $b$ (Matarrese et al. 1997; Moscardini et al. 1998):
\begin{equation}
\xi(r,z)=b^{2}(M_{\rm eff},z)\xi_{\rm DM}(r,z).
\label{eq:xi_obj}
\end{equation}
Here $M_{\rm eff}$ is the effective mass of the dark matter haloes in which the sources reside and $b$ is derived  according to the prescriptions of Sheth \& Tormen (1999).

Earlier analyses generally adopted the ${\mathcal N}(z)$ from Dunlop \& Peacock (1990) models, that were found to be inconsistent with the observational determination by Brookes et al. (2008). We have therefore repeated the analysis with the ${\mathcal N}(z)$ yielded by the present model that provides a good smooth fit of the latter data. As pointed out by Negrello et al. (2006), in the standard hierarchical clustering scenario, for $z\gsim 1$ angular scales $\theta\gsim 2^\circ$ correspond to linear scales where the correlation function is negative, so that the observation of a positive correlation function on large angular scales entails strong constraints on the bias parameter or, equivalently, on the effective halo mass, at high redshifts. This is because the amplitude of $w(\theta)$ on smaller scales, where the contributions from all the relevant redshifts are positive, requires the local value of the effective mass, $M_{\rm eff}(z=0)$, to be very large (Negrello et al. 2006 obtained $M_{\rm eff}(z=0) \simeq 10^{15}\,M_{\odot}$/h). For such values of $M_{\rm eff}$ the bias factor increases steeply with $z$, blowing up the negative contributions to $w(\theta)$. If $M_{\rm eff}$ is that large and almost independent of $z$, as it is found to be in the case of optically-selected QSOs (Porciani et al. 2004; Croom et al. 2004), $w(\theta)$ must be negative on scales above $2^\circ$--$3^\circ$, contrary to observations.

Although the new redshift distribution peaks at a somewhat lower $z$ than predicted by the pure luminosity evolution Dunlop \& Peacock (1990) model adopted by Negrello et al. (2006), and therefore the negative contributions to $w(\theta)$ are lower, we still find that $M_{\rm eff}$ cannot keep constant up to $z>1$. The observed $w(\theta)$ can be accounted for if $M_{\rm eff}$ decreases with increasing $z$ mirroring the evolution with cosmic time of the characteristic halo mass entering the non-linear regime. The best fit is obtained with a somewhat lower value for the local effective mass, $M_{\rm eff}(z=0) \simeq 10^{14.5}\,M_{\odot}$/h (Fig.~\ref{fig:wth}).

%%%%%%%%%%%%%%%%%%%%%%%%%%%%%%%%%%%%%%%%%%%%%%%%%%%%%%%%%%%%%%%%%%%%%%%%%%
\subsection{Integrated Sachs--Wolfe (ISW) effect}

The ISW effect describes the influence of the evolution of the gravitational potential in time-variable, linear, metric perturbations on CMB photons that traverse them. A non-zero effect arises if the gravitational potential decays, as in the case of an open universe when the effect of the space curvature is important, or when the dynamics of the universe is dominated by dark energy. As first pointed out by Crittenden \& Turok (1996) a promising way of probing the ISW effect is through correlating fluctuations in the Cosmic Microwave Background (CMB) with large-scale structure. Since the timescale for the decay of the potential is of the order of the present-day Hubble time, the effect is largely canceled on small scales, because photons travel through multiple density peaks and troughs. This is why surveys covering large areas of the sky and probing the large scale distribution up to $z\sim 1$ are necessary. These are the properties of the NVSS survey, which is therefore a particularly well-suited ISW probe, and indeed has been extensively exploited to look for the ISW signal (Boughn \& Crittenden 2004, 2005; Pietrobon et al. 2006; McEwen et al. 2007, 2008; Ho et al. 2008; Giannantonio et al. 2008; Raccanelli et al. 2008).

The comparison of the correlations inferred from the data with model predictions requires once again the redshift distribution and the bias parameter as a function of redshift. Again, all analyses carried out so far have used redshift distributions inconsistent with the CENSORS results, and this has motivated our re-analysis exploiting the redshift distribution yielded by the model. In addition we have redone the cross-correlation between the spatial distribution of NVSS sources brighter than 10 mJy and the CMB map from WMAP 5yr data provided by Delabrouille et al. (2009). We have also exploited the WMAP 5yr ILC map (Hinshaw et al. 2009) for cross-checks only, because it is declared to be sub-optimal for the angular scales we are interested in.

Since foreground cleaning of CMB maps is never perfect, we need to use a mask to remove the most contaminated areas. The choice of the mask must be done carefully. In fact, while a small mask allows a wider statistics to compute the CCF, it can also lead to a higher residual foreground contamination on the CMB map, thus diluting the CCF signal. We tried both the KQ85 and KQ75 masks provided by the WMAP team, which mask 18\% and 28\% of the sky respectively, including the strongest point sources in the WMAP channels.

\begin{figure}
\begin{center}
\includegraphics[totalheight=9cm, width=6cm, angle=90.]{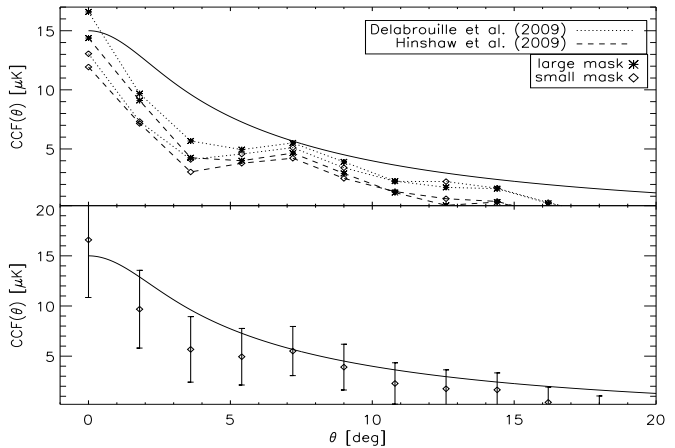}
\caption{WMAP--NVSS cross-correlation function (CCF). The upper panel compares CCF estimates obtained using different reconstructions of CMB maps from WMAP 5yr data (Delabrouille et al. 2009, dotted line; Hinshaw et al. 2009, dashed line) and different masks (KQ85, diamonds, and KQ75, asterisks; masks provided by the WMAP team). The solid line shows the expected signal for a concordance cosmology, based on the redshift distribution yielded by the present model and the bias function described in the text. In the lower panel, such signal is compared with our best estimate of the WMAP--NVSS CCF (CMB map by Delabrouille et al. 2009 and large mask; points with with $1\sigma$ error bars)} \label{fig:CCF}
\end{center}
\end{figure}

The estimate of the NVSS/CMB correlation, as a function of the angular scale, and of the associated errors, was performed as in Raccanelli et al. (2008), where the formalism used to compute the theoretical cross-correlation function is also presented.
The results for both CMB maps and masks are shown in the upper panel of Fig.~\ref{fig:CCF}. The CCFs computed using the extended KQ75 mask are significantly higher, suggesting that the KQ85 is not sufficient to avoid a significant effect of the residual contamination present in the CMB maps. Once the same mask is adopted, the curves corresponding to the different CMB maps are quite similar.

In the following, we used the Delabrouille et al. (2009) CMB map with the extended KQ75 mask. To estimate the errors on the empirical CCF we  simulated $\sim$ 3800 NVSS maps by randomly redistributing the unmasked pixels of the true NVSS map. For each simulated NVSS map we computed the CCF with the CMB map. In this way we obtained a model-independent test of the null-hypothesis (i.e. no ISW signal). Since the distribution of sources in the simulated maps does not trace the large scale density field with which the CMB interacts, the distribution of the simulated values of the CMB-NVSS cross-correlation function (CCF($\theta$)) allows us to estimate the level of significance at which the null hypothesis can be rejected. The results are shown in the lower panel of Fig.~\ref{fig:CCF}.
The CCF differs from zero at 2.8 $\sigma$ for the first point and at 2.4-1.7 $\sigma$ for the following 4 points.
Since the data points and their errors are correlated, the overall significance of the ISW signal cannot be straightforwardly derived from the significance of each point. Therefore, we have estimated it from the fraction of simulated maps yielding a CCF($\theta$) at the observed level, or higher. This fraction turned out to be $\sim 2.6\cdot 10^{-4}$, which implies a significance of the ISW signal one order of magnitude higher than found in Raccanelli et al. (2008) and in other previous analyses. This improvement is due to a better CMB map, as allowed by 5 years of integration by WMAP, and, even more, to the use of the wider KQ75 mask.

As shown by Fig.~\ref{fig:CCF}, lower panel, the cross-correlation power spectrum between the surface density fluctuations of NVSS sources and the CMB fluctuations expected for the ``concordance'' $\Lambda$CDM cosmology turns out to be in good agreement with the empirical determination. As in the case of $w(\theta)$ since we have now a reliable observational estimate of the redshift distribution of NVSS sources, the critical ingredient is the evolution with redshift of the bias factor $b(z)$ of radio sources. Without a reliable theory for it any strong inference from ISW estimates on the need of new gravitational physics on cosmological scales (Afshordi et al. 2008) or of a large local primordial non-Gaussianity (Afshordi \& Tolley 2008) is premature.

%
%\begin{figure}
%\begin{center}
%\includegraphics[totalheight=9cm, width=6cm, angle=90.]{ISW_cfr.eps}
%\caption{Estimates of the WMAP--NVSS cross-correlation function (CCF) with $1\sigma$ error bars. The solid line is the results obtained from the present model as in the previous figure. } \label{fig:CCF}
%\end{center}
%\end{figure}
%%%%%%%%%%%%%%%%%%%%%%%%%%%%%%%%%%%%%%%%%%%%%%%%%%%%%%%%%%%%%%%%%%%%%%%%%%
\section{Conclusions} \label{sec:Conclusions}

We have worked out a new analytical evolutionary model that describes the population properties of radio sources at frequencies $\lsim 5$ GHz, thus complementing the De Zotti et al. (2005) model, holding at higher frequencies. The model provides a satisfactory fit to a broad variety of data including luminosity functions of steep- and flat-spectrum sources, source counts at all frequencies from 0.15 to 5 GHz and over the full flux density range covered by observations, and redshift distributions.

We find that luminosity evolution is still sufficient to fit the wealth of available data on radio AGNs. However, the fit requires a luminosity-dependent decline of space densities of steep-spectrum sources at high redshifts, with positive evolution continuing up to higher redshifts for the more luminous sources, thus confirming earlier indications of a ``downsizing'' also for radio sources. The data do not require a similar behaviour for FSRQs.  However, the redshift at which the evolution peaks is found to be substantially higher for these sources than for the lower luminosity BL Lac objects, again consistent with  ``downsizing''. Note that, while earlier studies (Hook et al. 1998; Waddington et al. 2001; Vigotti et al. 2003; Cirasuolo et al. 2005, 2006) were based on limited data sets, the evidence presented here comes from a global analysis of all the available data. %The parameters providing the formal best fit imply a substantially stronger ``downsizing'' effect for steep- than for flat-spectrum sources. However, it must be noted that we have considered only one population of steep spectrum sources, but two populations of flat spectrum sources. The lower luminosity flat-spectrum population, BL Lac objects, was also found to evolve much less than the bright flat-spectrum radio quasars. Also, there are significant degeneracies among the model parameters, so that more data are necessary to assess the significance of difference between the evolutionary properties of the two populations.

The upturn of source counts at sub-mJy levels is accounted for by a straightforward extrapolation, exploiting the well established correlation between far-IR and radio luminosities, of evolutionary models accounting for the far-IR counts and redshift distributions of star-forming galaxies. As shown by earlier analyses (Yun et al. 2001; Best et al. 2005), a comparison of radio and $60\mu$m local luminosity functions indicates deviations from a linear relation at low luminosities. We present a simple analytic expression describing the correlation over the full luminosity range. The currently available surveys are not deep enough to provide an independent test of the deviations from linearity, but surveys with the new generation of radio telescopes (e.g. the SKA and its pathfinders) will.

The directly counted sources already account for most of the extragalactic background intensity determined by means of the classical T--T plot method. The dominant contribution comes from radio AGNs; the contribution of star-forming galaxies is higher at the lowest frequencies where it reaches 39.5\% percent.

We also discuss the implications of the new model for the interpretation of data on large-scale clustering of radio sources and on the Integrated Sachs-Wolfe effect, based primarily on the NVSS survey. Since both the sky coverage and the depth of this survey are much higher than those of large surveys in all the other wavebands, it provides unique constraints on the high-$z$ evolution of the bias factor, $b(z)$, on very large scales. The present analysis confirms and updates the conclusion of Negrello et al. (2006) that, if we adopt the Sheth \& Tormen (1999) formalism to derive $b(z)$ as a function of the effective halo mass, $M_{\rm eff}$, associated to radio sources, $M_{\rm eff}$ must decrease at $z > 1$ rather than keeping essentially constant as found for optically selected QSOs. In the absence of a reliable theory for the evolution of the bias parameter of radio sources, any claim on the need of new gravitational physics on cosmological scales or of a large local primordial non-Gaussianity, based on the spatial distribution of NVSS sources is, at best, premature.

%%%%%%%%%%%%%%%%%%%%%%%%%%%%%%%%%%%%%%%%%%%%%%%%%%%%%%%%%%%%%%%%%%%%%%%%%%
\section*{Acknowledgements}
We are grateful to the referee for thoughtful comments that helped substantially improving the paper. Partial financial support for this research has been provided to MM, AB, SR and GDZ, by the Italian ASI (contracts Planck LFI Activity of Phase E2 and I/016/07/0 `COFIS').

This research used resources of the National Energy Research Scientific Computing Center, which is supported by the Office of Science of the U.S. Department of Energy under Contract No. DE-AC02-05CH11231.

%%%%%%%%%%%%%%%%%%%%%%%%%%%%%%%%%%%%%%%%%%%%%%%%%%%%%%%%%%%%%%%%%%%%%%%%%%

\end{document}